\documentclass[showpacs,amsmath,amssymb]{revtex4}
\usepackage{epsfig}
\input{epsf}
\usepackage{amssymb}
\begin{document}
\title{Smooth dynamical (de)-phantomization of a scalar field in simple 
cosmological models}

\author{Alexander A. Andrianov}
\affiliation{Dipartimento di Fisica and INFN, Via Irnerio 46,40126 Bologna,
Italy\\
V.A. Fock Department of Theoretical Physics, Saint Petersburg State University,
198904, S.-Petersburg, Russia} 
\author{Francesco Cannata}
\affiliation{Dipartimento di Fisica and INFN, Via Irnerio 46,40126 Bologna,
Italy}
\author{Alexander Y. Kamenshchik}
\affiliation{Dipartimento di Fisica and INFN, Via Irnerio 46,40126 Bologna,
Italy\\
L.D. Landau Institute for Theoretical Physics of the Russian
Academy of Sciences, Kosygin str. 2, 119334 Moscow, Russia}

\begin{abstract} 
Simple scalar field cosmological models are considered 
describing gravity assisted  crossing of the phantom divide line. 
This crossing or (de)-phantomization characterized by the change of 
the sign of the kinetic term of the scalar field  
is smooth and driven 
dynamically by the Einstein equations. Different cosmological scenarios,
including the phantom phase of matter are sketched.  
\end{abstract}
 \maketitle

\section{Introduction}
The discovery of cosmic acceleration \cite{cosmic} has become 
one of the main challenges for modern theoretical physics,
because it has put forward the problem of the so called dark energy, 
responsible for this acceleration.
This dark energy should possess a negative pressure such that the relation
between pressure and energy density is less than $-1/3$ \cite{dark}.
A considerable amount of dark energy models was proposed 
(see, e.g. \cite{quintessence,darkmodel}), which now have to confront 
with growing quantity of observational data.    
  
Being enigmatic by itself, the phenomenon of the cosmic acceleration -
dark energy - negative pressure hides inside another subphenomenon,
which is not yet so well confirmed experimentally, but if 
confirmed, it would become, perhaps, more intriguing than all the 
preceeding cosmological puzzles. We mean here, the phenomenon of the 
so called phantom dark energy, i.e. the matter with the relation 
between the pressure and dark energy $k \equiv p/\rho$ 
less than $-1$ \cite{phantom}. 
Some observations indicate that the present day 
value of the parameter $k < -1$ provides the best fit.  
According to some authors, the analysis of observations, 
permits  
to specify the existence of the moment when the universe
changes the value of the parameter $k$ from that the region $k > -1$ to 
$k < -1$ \cite{phant-obs}. This transition is 
called ``the crossing of the phantom divide line''. Other authors
speak about double crossing of the phantom divide line (\cite{double}). 
In any case the existence of the phantom phase of cosmological 
evolution put forward some important theoretical problems.      
For example, considering a perfect fluid with barotropic equation 
of state, where $ k = const$ and $k < -1$ one can easily see that 
a universe during its evolution 
after a finite amount of time encounters a special kind 
of cosmological singularity, characterized by an infinite value of
the velocity of expansion, which has been dubbed ``Big Rip'' 
\cite{rip,star-rip}.   
Being a little bit unusual in comparison with well accepted Big Bang and
Big Crunch singularities, the Big Rip singularity is quite similar to them
from the formal (geometrical) point of view. Notice, that in the modern 
cosmological literature also other ``exotic'' types of singularities are
under consideration \cite{we-tach,Barrow-sudden, Shtanov, topor,stef}.

Another difficulty connected with the problem of phantom energy 
has a more fundamental character. It is well-known that the models 
with minimally coupled scalar field are very efficient both for the 
description of the inflationary stage of the development of the early universe
and for the explanation of the late-time cosmic acceleration
phenomenon \cite{quintessence}.    
However, it is easy to see, that the standard minimally coupled scalar 
field cannot give rise to the phantom dark energy, because in this model
the absolute value of  energy density is always greater than that of 
pressure, i.e. $|k| < 1$. A possible  way out of this situation 
is the consideration of the scalar field models with the negative 
kinetic term.  Again, being a little unusual from the point of view 
of common wisdom, these models have become quite popular novadays.
 
Perhaps, the most thrilling problem arising in connection with 
the phantom energy is  the crossing of the phantom 
divide line. The general belief is that this crossing is not admissible
in simple minimally coupled models and explanation of this 
phenomenon requieres  more complicated models such a multifield ones
or models with non-minimal coupling between scalar field and gravity 
(see e.g. \cite{B,divide,Periv,CD}).

In the present paper, we would like to explore the consequences of 
the possibility
of existence of simple minimally coupled scalar field models allowing 
the crossing of the phantom divide line.  
As a matter of fact, we would like to present some toy 
examples, suggesting that such a crossing could be unavoidable and 
dictated  by the continuity of the solution of Einstein equations. 
In the next section we discuss in some detail a simple toy model,
giving an illustration of the change of sign of the kinetic term 
in the scalar field cosmology.  
In the third section we sketch some other models and in the last section 
we attempt to discuss the above consideration in a more general context.

\section{Toy model with a crossing of the phantom divide line}
Let us consider a simple cosmological model, representing a flat Friedmann 
universe with a metric 
\begin{equation}
ds^2 = dt^2 - a^2(t)dl^2
\label{Friedmann}
\end{equation}
filled with a barotropic fluid with an equation of state 
\begin{equation}
p = k\rho,
\label{barotropic}
\end{equation}
where $\rho$ and $p$ are energy density and pressure of the fluid 
respectively and $-1< k \leq 1$. 
Choosing conveniently the normalisation of the constants in the theory
one can write the Friedmann equation in a simple form
\begin{equation}
h^2 = \rho,
\label{Friedmann1}
\end{equation}
where the Hubble variable is defined as $h \equiv \frac{\dot{a}}{a}$. 
The energy conservation law is 
\begin{equation}
\dot{\rho} = -3h(\rho + p)
\label{energy}
\end{equation}
which together with Eq. (\ref{barotropic}) gives immediately
\begin{equation}
\rho = \frac{\rho_0}{a^{3(1+k)}}.
\label{energy1}
\end{equation}
From Eqs. (\ref{energy1}) and (\ref{Friedmann1}) one gets
\begin{equation}
a = a_0 t^{\frac{2}{3(1+k)}},
\label{radius}
\end{equation}
or, in other terms,
\begin{equation}
h = \frac{2}{3(1+k)t}.
\label{Hubble}
\end{equation}
Another useful equation obtained by differentiation of the 
Eq. (\ref{Friedmann1}) and substitution into it Eq. (\ref{energy}) reads:
\begin{equation}
\dot{h} = -\frac32(\rho+p).
\label{Friedmann2}
\end{equation}
The pressure can be written down as 
\begin{equation}
p = -\frac23\dot{h} - h^2.
\label{pressure}
\end{equation}

It is well known that for a given cosmological evolution $h(t)$ 
satisfying some simple conditions 
one can find a minimally  coupled scalar field cosmological model 
with Lagrangian  
\begin{equation}
L = \frac{\dot{\phi}^2}{2} - V(\phi), 
\label{Lagrange}
\end{equation}
which contains this evolution as a particular solution, provided 
suitable initial conditions are chosen (see, e.g. \cite{Starob,we-tach} 
and references therein).
 
The energy density and pressure of the scalar field (\ref{Lagrange})
are 
\begin{equation}
\rho = \frac{\dot{\phi}^2}{2} + V(\phi),
\label{energy-scal}
\end{equation}
\begin{equation}
p = \frac{\dot{\phi}^2}{2} - V(\phi).
\label{pressure-scal}
\end{equation}
From Eqs. (\ref{energy-scal}), (\ref{pressure-scal}),
(\ref{Friedmann1}) and (\ref{pressure}) one has
\begin{equation}
\dot{\phi}^2 = (\rho + p) = -\frac23 \dot{h},
\label{scal}
\end{equation}
\begin{equation}
V(\phi) = \frac12(\rho-p) = \frac{\dot{h}}{3} + h^2.
\label{scal1}
\end{equation}
Equation (\ref{scal1}) gives the potential $V(t)$ as a function of $t$. 
Finding $\phi(t)$ by integrating Eq. (\ref{scal}), inverting this function
and substituting a function $t(\phi)$ into Eq. (\ref{scal1}), one finds the 
potential $V$ as a function of the scalar field $\phi$. Sometimes this can 
be done explicitly. For example, the evolution (\ref{Hubble}) could be 
reproduced in the extensively studied \cite{exp,we-tach}
cosmological model with an exponential potential 
\begin{equation}
V(\phi) \sim \exp\left(-\frac{3\sqrt{1+k}\phi}{2}\right).
\label{exponent}
\end{equation}

Apparently, the super-accelerated evolution (\ref{Hubble}) with 
$k < -1$ cannot be reproduced by means of the minimally coupled scalar field 
model (\ref{Lagrange}), but can be easily obtained in the framework of the 
model with phantom scalar field
\begin{equation}
L = -\frac{\dot{\phi}^2}{2} - V(\phi). 
\label{Lagrange1}
\end{equation}

Our starting point is the  remark 
made in \cite{Chervon,Yurov,Yurov1}, 
attracting the attention to some interesting 
features of the potential (\ref{scal1}) represented as a function of time $t$.
The point is that the volume function 
\begin{equation}
\psi(t) \equiv a^{3}(t)
\label{volume} 
\end{equation}
satisfies a simple second-order differential equation 
\begin{equation}
\ddot{\psi} = 9V(t)\psi,
\label{equation}
\end{equation}
which can be easily obtained from Eq. (\ref{scal1}). 
As has been already noticed above, the potential $V(t)$ is chosen 
in such a way to provide a given cosmological evolution.
For example, for the evolution (\ref{Hubble}) the form of the potential is
\begin{equation}
V(t) = \frac{2(1-k)}{9(1+k)^2t^2}.
\label{pot-time}
\end{equation}
For simple forms of the potential one can find the general 
solution of Eq. (\ref{equation}), which contains together with the solution 
used for the construction of this potential also another independent solution.
For the potential (\ref{pot-time}) the general solution looks like
\begin{eqnarray}
&&\psi(t) = \psi_1 t^{\alpha_1} + \psi_2 t^{\alpha_2},\nonumber \\
&&\alpha_1 = \frac{2}{1+k},\nonumber \\
&&\alpha_2 = \frac{k-1}{1+k},
\label{solution}
\end{eqnarray}
where $\psi_1$ and $\psi_2$ are nonnegative constants.  
(In papers \cite{Yurov,Yurov1} the solution (\ref{solution}) was written 
down for the case of dust $k = 0$.)  
Now, knowing $\psi(t)$ one can find $h(t) = \frac{\dot{\psi}}{3\psi}$
and $\dot{h}$. Substituting the obtained  $\dot{h}(t)$ into Eq.
(\ref{scal1}) one can try to find the dependence $\phi(t)$. Inversion of
this dependence $t(\phi)$ and its substitution into the expression 
for the potential as a function of time (\ref{pot-time}) would have given 
the potential $V(\phi)$, whose form is different from the exponential one 
(\ref{exponent}) when $\psi_2 \neq 0$. 

One can derive the equation for the functions $t(\phi)$ compatible with 
the potential $V(t)$ from Eq. (\ref{exponent}). First of all, notice that 
the field $\phi$ should satisfy the Klein-Gordon equation
\begin{equation}
\ddot{\phi} + 3h \dot{\phi} +\frac{dV}{d\phi} = 0.
\label{KG}
\end{equation}
Dividing Eq. (\ref{KG}) by $\dot{\phi}$ one has
\begin{equation}
\frac{\ddot{\phi}}{\dot{\phi}} + 3h + \frac{1}{\dot{\phi}}
\frac{dV}{d\phi} = 0.
\label{KG1}
\end{equation}
It is convenient to rewrite the second time derivative 
of the scalar field as
\begin{equation}
\ddot{\phi} = \dot{\phi} \frac{d\dot{\phi}}{d\phi} = 
\frac{1}{t'}\left(\frac{1}{t'}\right)',
\label{second}
\end{equation}
where we have used the relation
\begin{equation}
t' = \frac{1}{\dot{\phi}},
\label{prime}
\end{equation}
with the ``prime'' denoting the differentiation with respect to 
the scalar field $\phi$. 
Now, we would like to take the time derivative of Eq. (\ref{KG1}).
The first term gives
\begin{equation}
\frac{d}{dt}\left(\frac{\ddot{\phi}}{\dot{\phi}}\right) = \frac{1}{t'}
\left(\frac{1}{t'}\right)''.
\label{firstterm}
\end{equation}
Then, using the relations (\ref{scal1}) and (\ref{prime})  one gets
\begin{equation}
3\dot{h} = -\frac{9}{2t^{'2}}.
\label{secondterm}
\end{equation}
Then we make the following transformation:
\begin{equation}
\frac{1}{\dot{\phi}}\frac{dV}{d\phi} = t^{'2}\dot{V}.
\label{thirdterm}
\end{equation}
The time derivative of the expression in the left hand side of 
Eq. (\ref{thirdterm}) is 
\begin{equation}
\frac{d}{dt}\left(\frac{1}{\dot{\phi}}\frac{dV}{d\phi}\right) =
\frac{1}{t'}(t^{'2}\dot{V})'.
\label{thirdterm1}
\end{equation}
Combining  expressions (\ref{firstterm}),(\ref{secondterm}) and 
(\ref{thirdterm1}), using the explicit form of the potential 
(\ref{pot-time}) and multiplying the resulting equation by $t'$ 
one arrives to the following equation:
\begin{equation}
\left(\frac{1}{t'}\right)'' -\frac{9}{2t'} -\frac{4(1-k)}{9(1+k)^2}
\left(\frac{t^{'2}}{t^3}\right)' = 0.
\label{nonlinear}
\end{equation}

Particular solutions of this equation being substituted into 
Eq. (\ref{pot-time}) generate different potentials $V(\phi)$ 
corresponding to different cosmological evolutions (\ref{solution}). 
It is easy to check that the solution 
\begin{equation}
t = \exp\frac{3\sqrt{1+k}\phi}{2},
\label{partic}
\end{equation}
corresponding to the potential (\ref{exponent}) and to the evolution 
with $\psi_2 = 0$, satisfies Eq. (\ref{nonlinear}).     
Instead of looking for other exact solutions (which is not an easy task)
we concentrate now on the qualitative study of Eq. (\ref{nonlinear}) 
or its equivalents. 

First of all, let us notice that the solution (\ref{solution}) tacitly assumes 
that the time parameter runs from $0$ to $+\infty$. When $t \rightarrow 
+\infty$ one can neglect the influence of the second term in the 
solution (\ref{solution}). Thus, the potential $V(\phi)$ in this limit 
has the asymptotic form (\ref{exponent}).
It is possible to calculate also the subleading term. 
The form of the Hubble variable corresponding to the evolution 
(\ref{solution}) is 
\begin{equation}
h = \frac{\alpha_1\psi_1 t^{\alpha_1-1} + 
\alpha_2\psi_2 t^{\alpha_2-1}}
{3(\psi_1 t^{\alpha_1} + 
\psi_2 t^{\alpha_2})},
\label{Hubble1}
\end{equation}
while its time derivative is 
\begin{equation}
\dot{h} = -\frac{\alpha_1\psi_1^2 t^{2\alpha_1} + 
\alpha_2\psi_2^2 t^{2\alpha_2} + (1-(\alpha_1-\alpha_2)^2)\psi_1
\psi_2 t^{\alpha_1+\alpha_2}}{3t^2(\psi_1 t^{\alpha_1} + 
\psi_2 t^{\alpha_2})^2}.
\label{Hubble3}
\end{equation}
At large values of $t,  \dot{h}$ written down up to 
the first non-leading term reads:
\begin{equation}
\dot{h} \approx -\frac{2}{3(1+k)t^2}\left(1 + \frac{2(k-3)\psi_2}
{(1+k)\psi_1}t^{\frac{k-3}{1+k}}\right).
\label{Hubble4}
\end{equation}
Substituting the expression (\ref{Hubble4}) into Eq. (\ref{scal1}) one finds
the subleading correction to the scalar field $\phi$ as a function 
of $t$. Inverting this relation and inserting the result into
Eq. (\ref{pot-time}) we obtain the potential
$V(\phi)$ including the first nonleading term:
\begin{equation} 
V(\phi) \sim \exp\left(\frac{-3\sqrt{1+k}\phi}{2}\right)
\left(1+3\frac{\psi_2}{\psi_1}
\frac{k-3}{\sqrt{1+k}}\exp\frac{3(k-3)\phi}{2\sqrt{1+k}}
\right).
\label{subleading}
\end{equation}
The above formula can be elucidated as follows:
if one consider the cosmological evolution described by Eq. (\ref{solution})
with non-zero value of the parameter $\psi_2$, one can state that 
there exists 
a potential describing this evolution. In spite of the absence of 
an explicit expression for this potential we know that its asymptotic 
behaviour is described by the formula (\ref{subleading}). 

Now, let us consider the   evolution of a cosmological model, 
which is characterized by Eq. (\ref{Hubble1}) and the potential whose 
asymptotic form is given by Eq. (\ref{subleading}). It is easy to show
that at some moment $t_0$, such as $0 < t_0 < \infty$, the time derivative 
of the Hubble variable vanished. Indeed, a simple analysis 
of the formula
(\ref{Hubble3}) gives this value 
\begin{equation}
t_0 =  \left(\frac{(3-k)\sqrt{2(1-k)}+4(1-k)}{2(1+k)}
\frac{\psi_2}{\psi_1}
\right)^{\frac{1+k}{3-k}}.
\label{time-phantom}
\end{equation}
Obviously,  $t_0$ exists for any value of $\psi_2 \neq 0$. 
The evolution of the model given by Eq. (\ref{Hubble1}) is consistently 
defined at $t > t_0$ in the standard scalar model (\ref{Lagrange}) 
with a potential whose asymptotic form at large values of $t$ was 
presented above (\ref{subleading}). (The consistency is based on the 
fact that at $t > t_0$ the time derivative of the Hubble constant 
is always negative). It is possible also to find the approximate expression
for the potential $V(\phi)$ for $t \rightarrow t_0+$. 
it is well known that the Klein-Gordon and Friedmann equations allow a shift
of the scalar field by some constant. It is convenient to fix the 
scalar field as $\phi \rightarrow 0$ when   
$t \rightarrow t_0+$. The time derivative of the Hubble variable at
$t \rightarrow t_0+$ behaves as 
\begin{equation}
\dot{h} = -H(t-t_0),
\label{Hubble5}
\end{equation}
where 
\begin{equation}
H = \frac{\sqrt{8(1-k)}(3-k)^2\psi_1\psi_2t_0^{-\frac{4k}{1+k}}}
{(1+k)^3\left(\psi_1 t_0^{\frac{2}{1+k}} + \psi_2 t_0^{\frac{k-1}{k+1}}
\right)^2}.
\label{second-der}
\end{equation}
Substituting the expression (\ref{Hubble5}) into Eq. (\ref{scal1}) 
one can integrate the latter, getting
\begin{equation}
\phi \approx \sqrt{\frac{8H}{27}}(t-t_0)^{3/2},
\label{scal-zero}
\end{equation}
whose inversion gives
\begin{equation}
t = t_0 +\left(\sqrt{\frac{27}{8H}}\phi\right)^{2/3}.
\label{scal-zero1}
\end{equation}
Thus, the potential results
\begin{equation}
V(\phi) = 
\frac{2(1-k)}{9(1+k)^2\left(t_0+
\left(\sqrt{\frac{27}{8H}}\phi\right)^{2/3}\right)^2}.
\label{pot-zero}
\end{equation}
It is rather straightforward to verify that the expressions for 
$\phi,\dot{\phi}$ and the potential $V(\phi)$ are regular when 
$t \rightarrow t_0+$ while the derivative $\frac{dV(\phi)}{d\phi}$ 
and $\ddot{\phi}$ are singular.   

Now, before discussing the subtle question of the behavior of the 
cosmological model under consideration at  $t = t_0$ and of 
the possibility of crossing  this point, we shall study briefly 
another branch of the evolution (\ref{Hubble1}): namely, that where
$t < t_0$.
For small values of $t$ to reproduce the evolution (\ref{solution})
one should use the phantom Lagrangian (\ref{Lagrange1}). Proceeding 
in the similar way one can find the corresponding potential 
of the phantom field in the subleading approximation:
\begin{equation}
V(\phi) =  
\frac{2(1-k)}{9(1+k)^2}\exp \left(-3\sqrt{\frac{2(1+k)}{1-k}}\phi
\right)\left((1+\frac{4\psi_1(1+k)}{\psi_2(3-k)}\exp\left(3\phi(3-k)\sqrt
{\frac{1}{2(1-k^2}}\right)\right).
\label{pot-zero1}
\end{equation}

One can easily describe what is going on in the model characterized by 
a negative kinetic term and the potential whose asymptotic form is given
by (\ref{pot-zero1}) in the vicinity of the point $t \rightarrow t_0-$. 

Now, we are in a position to scrutinize the possibility of matching 
 these two branches of the evolution (\ref{solution}) 
at $t < t_0$ and $t > t_0$. Are they really incompatible? 
Let us consider  a scalar field model with a negative kinetic term 
and a potential whose asymptotic form is that of Eq. (\ref{pot-zero1}).
Further, let us suppose that at some moment $t < t_0$ one has initial 
conditions on the values of $h,\phi$ and $\dot{\phi}$ which provide the 
evolution (\ref{solution}) where the values of the parameters are consistent
with those of the potential $V(\phi)$.  
Then approaching the time moment $t \rightarrow t_0-$ one arrives to the 
regime when $\dot{h}, \phi$ and $\dot{\phi}$ tend to vanish. Nevertheless,
all the geometric characteristics of the spacetime remain well defined 
(due to 
homogeneity and isotropy of the Friedmann cosmology all the curvature 
invariants are expressible through the cosmological factor $a$ and its
time derivatives, which are obviously finite). In contradistinction, 
the second time derivative of the scalar field at $t = t_0$ and 
$\frac{dV(\phi)}{d\phi}$ diverge, but this divergence is an integrable one.
This offers us an opportunity (and, perhaps, necessity) of a continuation 
of the  spacetime geometry and field configurations beyound this 
``divide line''. Clearly, such a smooth dynamical continuation which respects 
Einstein equations, entails the change of the sign of the kinetic term 
and the transition to the regime for $t > t_0$ described above.  

\section{Different regimes of crossing of phantom divide line}
The cosmological evolution presented in the preceding section 
describes the following scenario: the universe begins its evolution 
from the cosmological singularity of the ``anti-Big Rip'' type and 
its squeezing is driven by a scalar field with a negative kinetic term.
Then at the moment (easily calculated from Eq. (\ref{Hubble1}))
\begin{equation}
t_1 = \left(\frac{(1-k)\psi_1}{2\psi_2}\right)^{\frac{3-k}{1+k}}
\label{stop}
\end{equation} 
the contraction of the universe is replaced by an expansion.
At the moment $t = t_0 > t_1$ (see Eq. (\ref{time-phantom}) the kinetic 
term of the scalar field changes sign. Then, with the time growing 
the universe undergoes an infinite  power-law expansion. 
The graphic of the time dependence of $h(t)$ is presented in Fig. 1. 
\begin{figure}[h]
\epsfxsize 6cm 
\epsfbox{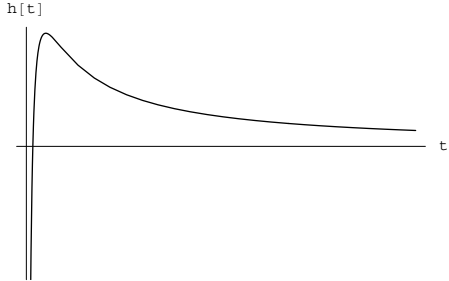}
\caption{$h(t)$ dependence in our toy model}
\label{Fig.1}
\end{figure}
Changing the sign of time parameter in all the equations  
the range $-\infty < t < 0$ one can consider the cosmological evolution 
running from $t = -\infty$ to $t = 0$. 
It  begins from the non-singular contraction and ends in 
a ``standard'' Big Rip singularity.
In this case at the moment $t = -t_0$ the scalar 
field undergoes the ``phantomization'', while the transition at the 
moment $t = t_0$, considered in the preceeding section can be called 
``dephantomization''. 

The toy model discussed above prrovides an illustration of these phenomena 
of phantomization and dephantomization, but from observational point of 
view, it would be more interesting to get an evolution beginning from 
the Big Bang and ending in the Big Rip singularity, after undergoing 
a phantomization transition. Instead of trying to construct some 
potential and cosmological evolution describing such a process we shall 
limit ourselves to giving a graphical presentation of the $h(t)$ -dependence
which could be responsible for such a scenario (see Fig. 2).  
This picture  represents the function $h(t) = \frac{A_0}{t(t_R -t)}$.
Here the moment $t = t_R$ is the moment of Big Rip, while the the moment
of phantomization is $t_0 = t_R/2$. In principle, the function $h(t)$ 
determines the model completely, including the form of the potential, even 
if when it is not written explicitly.   
\begin{figure}[h]
\epsfxsize 6cm 
\epsfbox{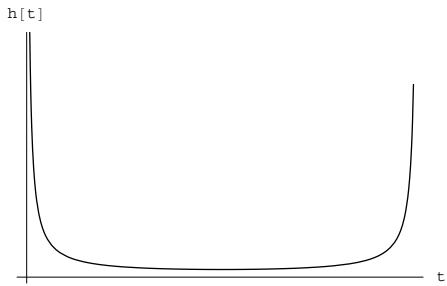}
\caption{$h(t)$ dependence in the model, describing the Big Rip.}
\label{Fig.2}
\end{figure}

As has been already mentioned in Introduction, some observations 
favor  double crossing of the phantom divide line \cite{double}.
The corresponding graphic $h(t)$ is presented in Fig. 3. 
\begin{figure}[h]
\epsfxsize 6cm 
\epsfbox{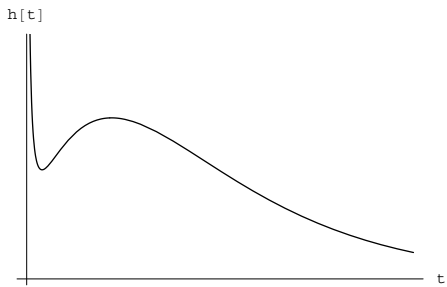}
\caption{$h(t)$ dependence in the model, describing a double crossing of the 
phantom divide line.}
\label{Fig.3}
\end{figure}
Also to this 
time dependence of $h(t)$ one can associate some potential and the evolution, 
in course of which the scalar field undergoes two smooth transitions:
phantomization at the point of mimimum of the curve $h(t)$ and the subsequent
dephantomization at the point of its maximum. 

It is suggestive to remark that the diagram Fig.3 reminds topologically 
the well-known van der Waals isoterm curve. The part of the 
van der Waals curve situated between the maximum and minimum, describing 
a metastable state (of supercool vapor or superheated liquid),
is analogous to the phantom phase in our Fig. 3.     
Models with double crossing of the phantom divide line
were considered also in \cite{Yurov}. The idea in \cite{Yurov} consisted in 
cutting off the phantom piece of the evolution with the subsequent 
sewing of remaining branches of the trajectory removes the problems 
connected with the crossing of the phantom divide line. In other terms,
the trajectories, considered in \cite{Yurov} correspond graphically to  $h(t)$ 
which instead of one maximum and one minimum (see Fig. 3) have one 
inflection point. Continuing the analogy with the van der Waals isoterms,
one sees that this curves corresponds to the critical van der Waals curve.  
Surely, such models can be studied, but their existence
does not exclude the existence of models described by Fig. 3 and in a frame
of a concrete model with fixed Lagrangian and initial conditions these 
two types of trajectories cannot be transformed one into other.

\section{Discussion} 
Above we have discussed the properties of some simple 
cosmological models based on minimally coupled scalar fields, 
which could, in our opinion, describe the crossing of the phantom divide
line, changing the sign of the kinetic term in the corresponding Lagrangian.   
As a starting point of our consideration we have used the observation 
\cite{Yurov,Yurov1} that having a simple form of the scalar field potential
, written as a function of time $V(t)$, one can find the general solution 
of the Einstein (Friedmann) equations for the Hubble variable $h(t)$
(see Eq. (\ref{solution}). 
Different particular solutions describing the evolution  of the Hubble 
variable 
correspond to different potentials as functions of the scalar field 
$V(\phi)$. All these solutions (excluding the obvious, when $\psi_2 =0$)
contain the time moment when the time derivative of the Hubble variable 
vanishes (see Eq. (\ref{time-phantom})) and the only consistent way 
of passing of this point respecting the continuity of the Einstein equations 
is the change of the sign of the kinetic term for the scalar field. 
This change can be called (de)-phantomization of the field.

All above are of technical nature and are confirmed 
by a direct analysis of the corresponding differential equations.
As far as physical interpretation is concerned one has two alternatives.
One can say that the sign of the kinetic term should be fixed a priori
(see, e.g. \cite{Periv})) and, hence the models, which imply the change 
of this sign should be discarded as non-physical ones. 
Here, we attempted to treat another alternative: i.e. to take this theory 
seriously because from mathematical point of view it looks  
consistent. In choosing this alternative we were guided by a belief 
that Einstein equations are more fundamental then the concrete form 
of the action for other fields.    
The idea about the dominant role of Einstein equations goes back to 
the classical works by Einstein, Infeld and Hoffmann \cite{EIH,EI}, where
it was shown that the motions of mass points (geodesics law) are determined 
by the Einstein equations for the gravitational field. Later this approach 
was confirmed in works by Lanczos \cite{Lanczos}, Fock \cite{Fock} and others.
In this respect the general relativity is quite different from  
electrodynamics, where the Lorentz's force law is quite independent 
of the Maxwell field equations (for a discussion see e.g. \cite{Bergmann}).
The derivability of the equation of motion for particles is based on 
the fact that the Einstein equations are non-linear and also they are        
subject to additional identities, namely the Bianchi identities which reduce 
the number of independent equations. These Bianchi identities imply 
conservation laws for  matter, present in the model under 
consideration. 

In fact, the argumentation, based on the Bianchi identities could 
be used also for getting information about the possible field configurations
for non-gravitational fields. Indeed, the Einstein equations for fields,
minimally coupled to gravity, have the form
\begin{equation}
R_{\nu}^{\mu} - \frac12 \delta_{\nu}^{\mu} R = \kappa T_{\nu}^{\mu}.
\label{Einstein}
\end{equation}
The Bianchi identities
\begin{equation}
\nabla_{\mu}\left(R_{\nu}^{\mu} - \frac12 \delta_{\nu}^{\mu} R\right) = 0
\label{Bianchi}
\end{equation}
imply some kind of the energy-momentum conservation law:
\begin{equation}
\nabla_{\mu}T_{\nu}^{\mu} = 0.
\label{conserv}
\end{equation}
For the case of the minimally coupled spatially homogeneous time-dependent 
scalar field with the standard sign 
of the kinetic term Eq. (\ref{conserv})   
is reduced to
\begin{equation}
\left(\ddot{\phi} + 3h\dot{\phi} + \frac{dV}{d\phi}\right)\dot{\phi} = 0,
\label{conserv1}
\end{equation}
which is equivalent to the Klein-Gordon equation when $\dot{\phi} \neq 0$.
For the phantom scalar field one has similarly
\begin{equation}
\left(-\ddot{\phi} - 3h\dot{\phi} + \frac{dV}{d\phi}\right)\dot{\phi} = 0.
\label{conserv2}
\end{equation}
The comparison between Eqs. (\ref{conserv1}) and (\ref{conserv2}) points 
to the opportunity of the change of type of the Klein-Gordon equation at 
the moment when $\dot{\phi} = 0$. In other words, while the Einstein 
equations require the change of the type of the scalar field Lagrangian
provided some form of the scalar field potential and the cosmological 
evolution are chosen, the Bianchi identities show why this transformation 
is possible.   

Before concluding, let us notice that we have shown that simple scalar 
field models driving the universe through the phantom divide line do 
exist. However, the natural question arises: given the potential as 
a function of the scalar field (not of the time) how general are 
the initial conditions providing the reaching of the point 
$\dot{\phi} = 0$. In other words, do we need the fine tuning 
(cf. \cite{CD}) ?. The answer to this question is not simple 
technically and, perhaps, is worth further investigations. 
Another interesting question is concerned with an opportunity of using 
some kind of modified local Lagrangian formalism for the description 
of smooth dynamical transition from one form kinetic term to another one.  

\section*{Acknowledgments}
The authors are grateful to A.A. Starobinsky for a useful discussion.
A.A. was partially supported by RFBR, grant No. 05-02-17477 and 
by the Programme UR, grant No. 02.01.299. 
A.K. was  partially supported by  RFBR, grant No. 05-02-17450and  
by the Research
Programme "Astronomy" of the RAS.

\end{document}